\documentclass{ws-ijmpcs}

\begin{document}

\markboth{Li Dengjie}{Cosmic Muon Induced Backgrounds in the DayaBay}

%
\catchline{}{}{}{}{}
%

\title{Cosmic Muon Induced Backgrounds in the Daya Bay Reactor Neutrino Experiment}

\author{Li Dengjie}

\address{Department of Modern Physics, University of Science and Technology of China, Hefei 230026, P.R.China\\
lidengjie@ihep.ac.cn}

\maketitle

\begin{history}
\received{Day Month Year}
\revised{Day Month Year}
\end{history}

\begin{abstract}
 Muon induced neutrons and long-lived radioactive isotopes are important background sources for low-energy underground experiments. We study the produced processes and properties of cosmic muon induced backgrounds, show the muon veto system used for rejecting these backgrounds and the methods to estimate residual backgrounds in the Daya Bay Reactor Neutrino Experiment.
\end{abstract}

\ccode{PACS numbers: 25.30.Mr,29.40.-n}

\section{Introduction}
Muons are secondary products of collisions between high-energy cosmic rays and atmosphere nuclei. The muon flux at sea level is roughly 160 $Hz/m^2$. Muon induced neutrons and long-lived radioactive isotopes are important background sources for low-energy underground experiments. There are three important muon induced backgrounds in the Daya Bay experiment , accidentals ,fast neutron and  $^8He/^9Li$.
\section{Muon Induced Backgrounds}
 Following are the processes that muon induces backgrounds.Muon induced neutron can be produced by
muon spallation that muon interactions with nuclei via a virtual photon producing a nuclear disintegration,
muon elastic scattering with neutrons bound in nuclei,
photo-nuclear reactions associated with electromagnetic showers generated by muons,
and secondary neutron production following any of the above processes.
Muon induced isotopes can be produced by muon spallation and their secondary shower particles\cite{WYF}.
\section{Muon Veto System}
\begin{figure}[!htb]
\begin{minipage}[b]{0.49\textwidth}
\centerline{\includegraphics[width=4.7cm]{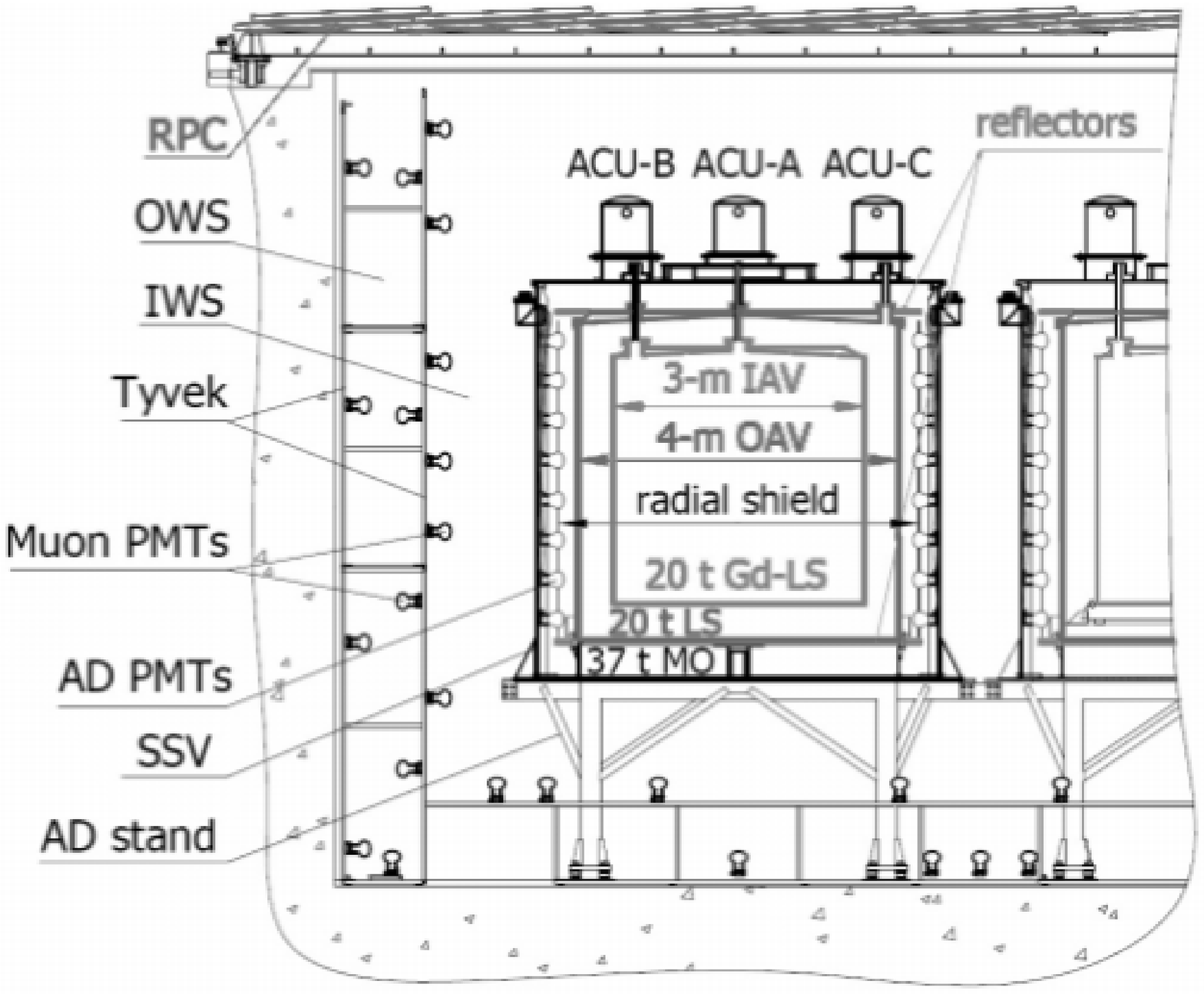}}
\vspace*{8pt}
\caption{Muon veto system \label{muon_veto_sys}}
\end{minipage}
\begin{minipage}[b]{0.49\textwidth}
\centerline{\includegraphics[width=4.7cm]{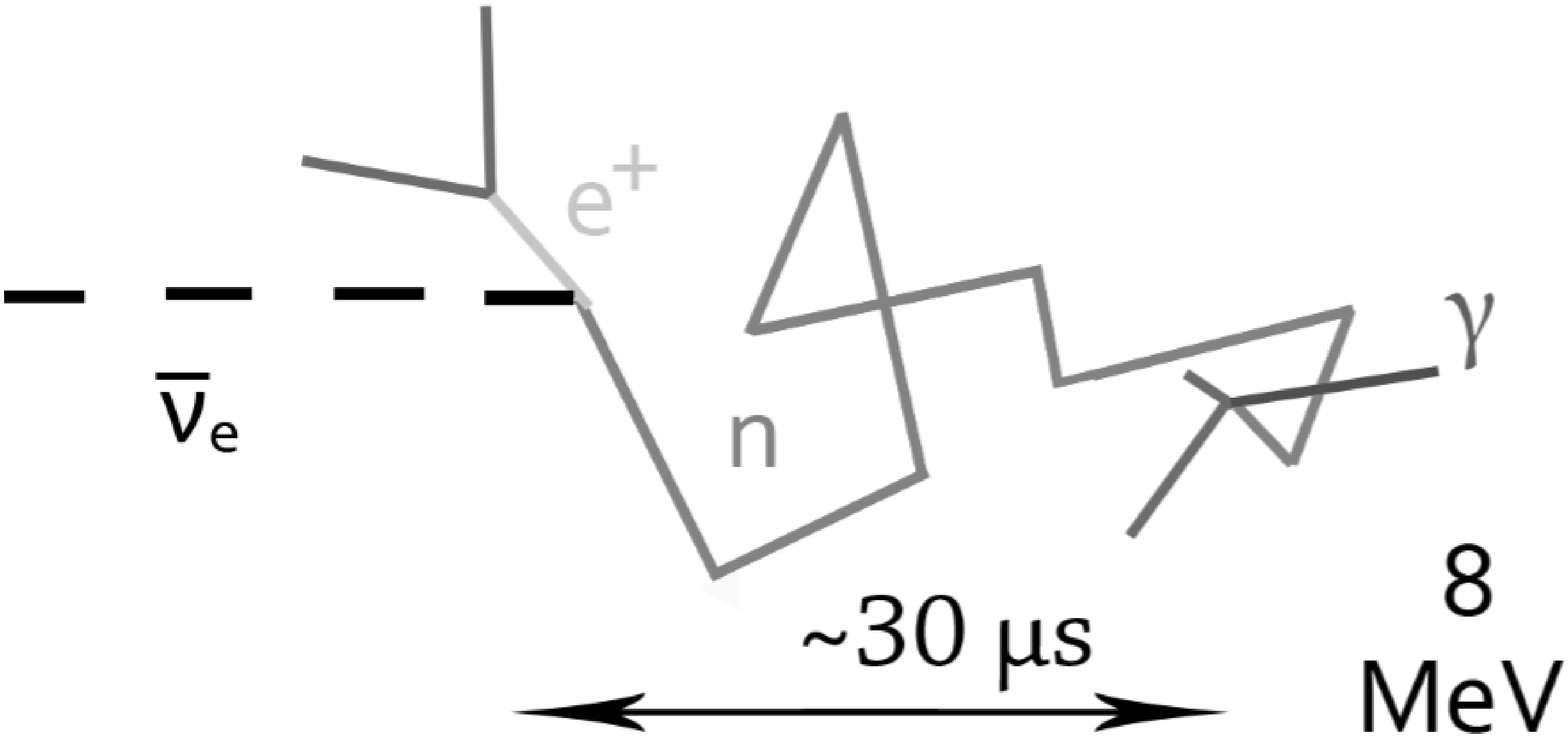}}
\vspace*{36pt}
\caption{Physical process of inverse beta decay  \label{ibd_pro}}
\end{minipage}
\end{figure}

Locating the detectors at sites with adequate overburden is the only way to reduce the muon flux and the associated background to a tolerable level. Being supplemented with a good muon identifier outside the detector, we can tag the muons going through or near the detector modules and reject backgrounds efficiently. Fig.\ref{muon_veto_sys} is the muon veto system in Daya Bay Reactor Neutrino Experiment\cite{PRL}.

This water buffer is used as a cherenkov detector to detect muons. Thus neutrons produced by muons in the detector module , the water buffer or surrounding rock will be effectively attenuated by the 2.5m water buffer. Together with RPC outside the water buffer, the combined muon tag efficiency is 99.5\%, with an uncertainty smaller than 0.25\%.
Veto time windows after tagged muon are [ 0 $\mu$s , 600 $\mu$s ] for water pool tagged muon,[ 0 ms , 1 ms ] for AD tagged muon,[ 0 s , 1 s ] for AD tagged shower muon,respectively.
\section{Three Backgrouds In DayaBay}
\subsection{Inverse beta decay}
The reaction employed to detect the $\overline{\nu}_e$ from reactor is the inverse beta decay.
\begin{equation*}
\overline{\nu}_e + p \to e^+ + n
\end{equation*}
Fig.\ref{ibd_pro} is the physical process of inverse beta decay. The reaction has a distinct signature : a prompt positron followed by a neutron-capture. Selections of  inverse beta decay events(IBD) are
\begin{romanlist}[(ii)]
\item Energy of prompt signal : [ 0.7 MeV , 12.0 MeV ],
\item Energy of delayed signal : [ 6.0 MeV , 12.0 MeV ],
\item Delta time between delayed and prompt signal: [ 1 $\mu$s , 200 $\mu$s ],
\item Multiple cut : no other $>$0.7 MeV triggers in 200 $\mu$s time window before prompt signal or after delayed signal.
\end{romanlist}

\subsection{Accidentals}
A single neutron capture signal has some probability to fall accidentally within the time window of a preceding signal due to natural radioactivity (U, Th, K, Co, Rn, Kr ), producing an accidental background. Some other long-lived cosmogenic isotopes, such as $^{12}B/^{12}N$, can fake the delayed 'neutron' signal if they have beta decay energy in the 6 MeV $\sim$ 12 MeV range.

The expected accidental background can be calculated as:
\begin{equation}
N_{accBkg}=\Sigma N_{n-like\ singles}^{i}\cdot(1-e^{R^i_{e^+-like\ triggers}\cdot 200\mu s}).
\label{accfunc}
\end{equation}
where, $N_{n-like\ singles}^{i}$ is rate of the triggers in prompt signal energy region that calculated every 4 hours , $R^i_{e^+-like\ triggers}$ is number of the triggers in delayed signal energy region that counted every 4 hours.
\subsection{Fast neutron}
\begin{figure}[!htb]
\begin{minipage}[b]{0.49\textwidth}
\centerline{\includegraphics[width=4.7cm]{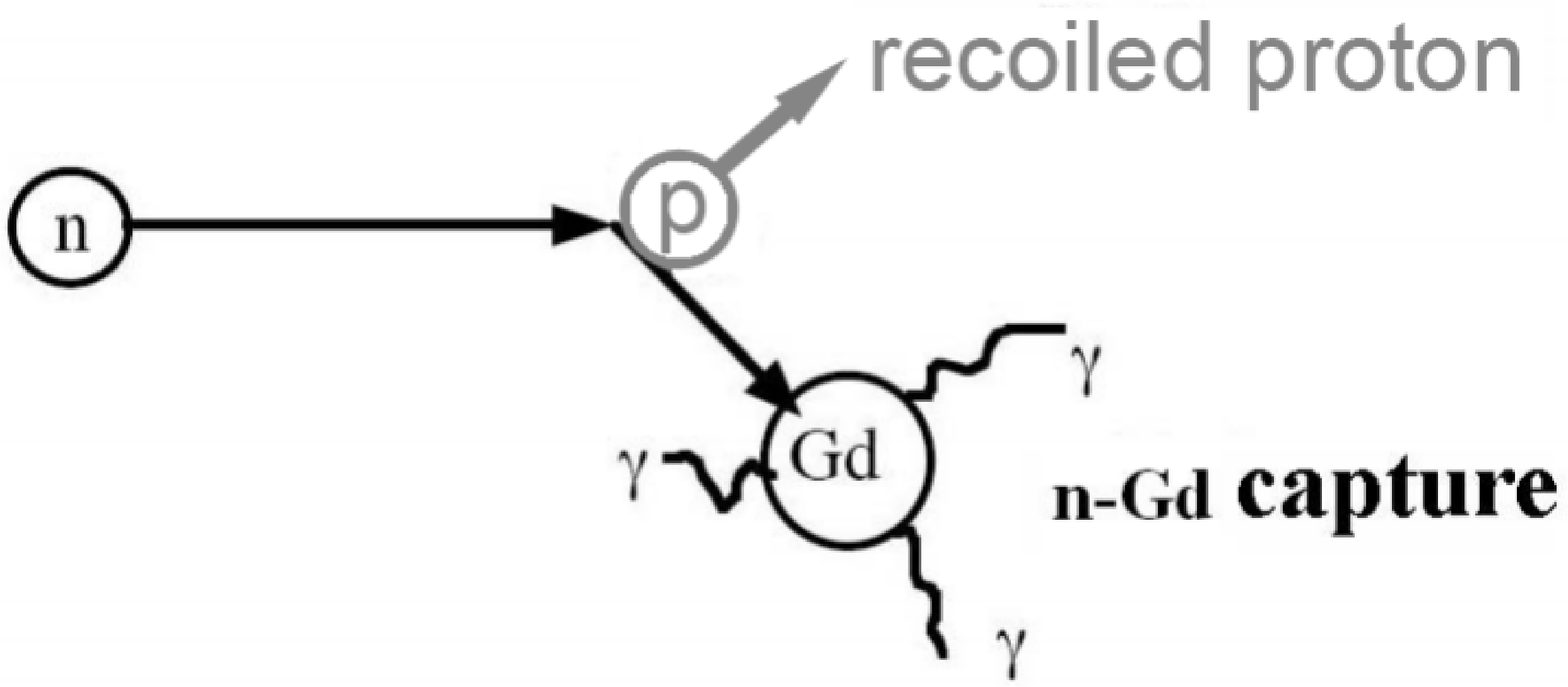}}
\vspace*{28pt}
\caption{Physical process of fast neutron \label{fn_pro}}
\end{minipage}
\begin{minipage}[b]{0.49\textwidth}
\centerline{\includegraphics[width=5cm]{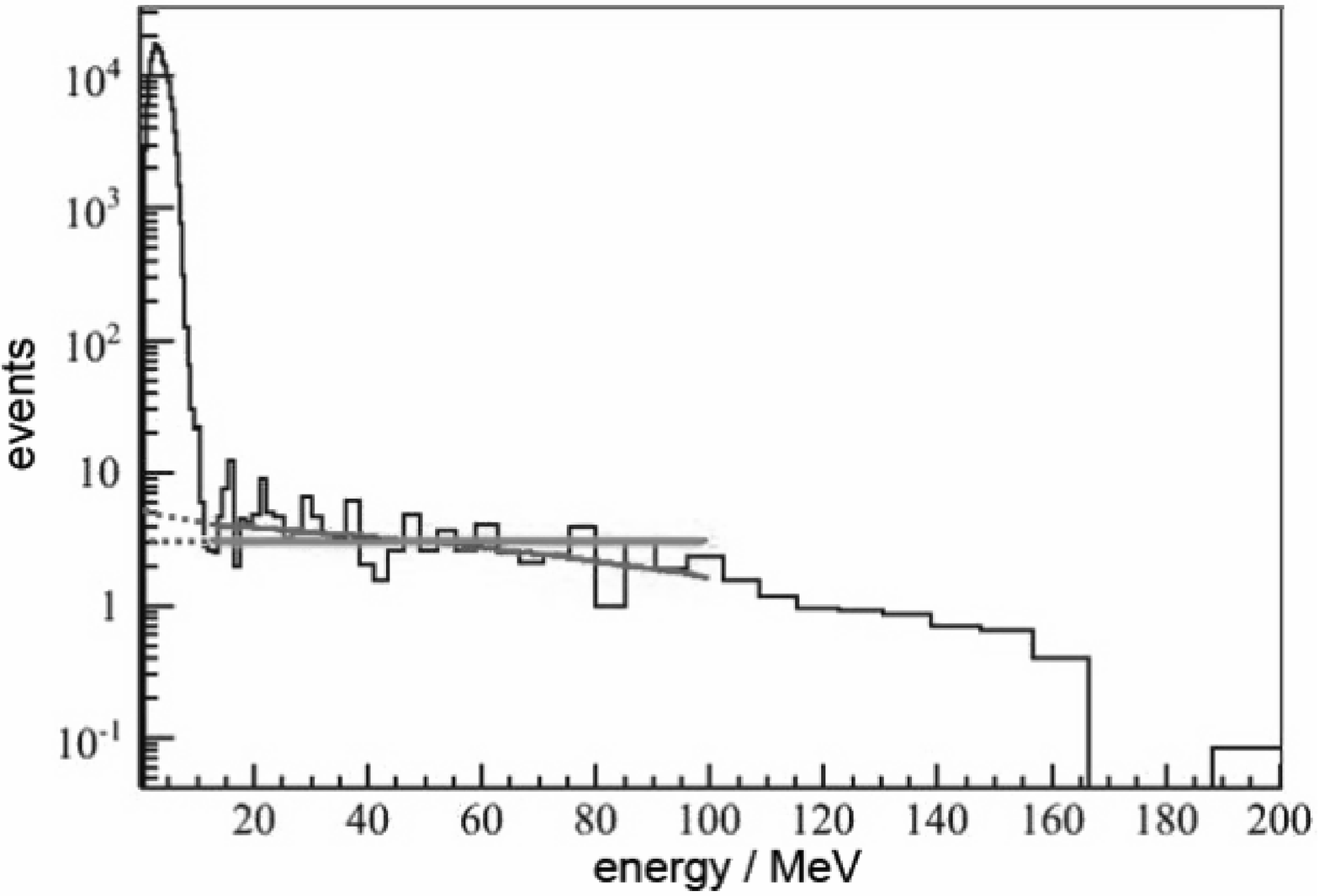}}
\vspace*{8pt}
\caption{Fitting method of fast neutron \label{fn_fit}}
\end{minipage}
\end{figure}

A fast neutron produced by cosmic muons in the surrounding rock or the detector can produce a signal mimicking the inverse beta decay reaction in the detector. Fig.\ref{fn_pro} is the physical process of fast neutron, the recoil proton generates the prompt signal and the capture of the thermalized neutron provides the delayed signal.The energy spectrum of prompt signal of only inner water pool tagged fast neutron between 0 MeV and 100 MeV is linear distribution.

Thus we use linear extrapolation to estimate fast neutron background (see Fig.\ref{fn_fit}). First,relax the 12 MeV prompt energy criterion in the IBD selection to 100 MeV, then take the mean of zero and first order polynomial fit values in [ 12 MeV , 100 MeV ] range to extrapolate the backgrounds into the [ 0.7 MeV , 12 MeV ] region.
\subsection{$^8He/^9Li$}
\begin{figure}[!htb]
\begin{minipage}[b]{0.49\linewidth}
\centerline{\includegraphics[width=4.7cm]{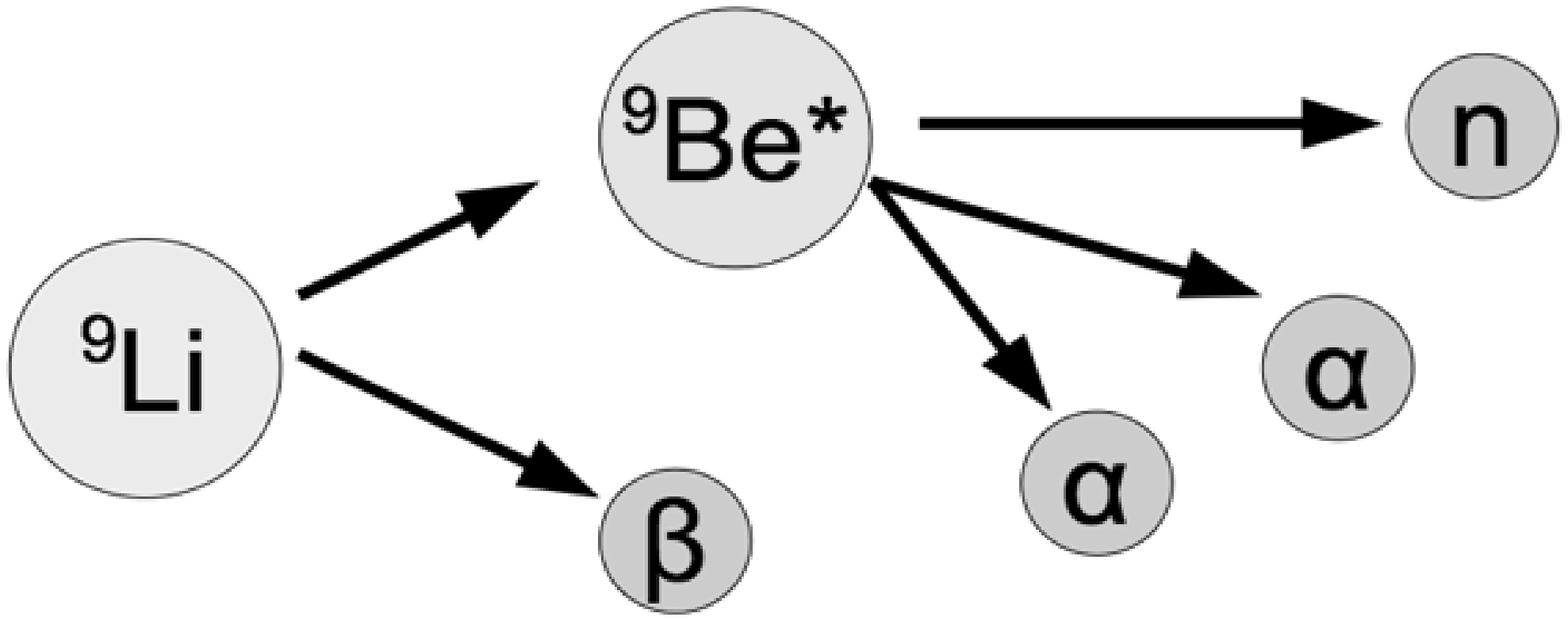}}
\vspace*{28pt}
\caption{Physical process of $^9Li$ \label{li_pro}}
\end{minipage}
\begin{minipage}[b]{0.49\linewidth}
\centerline{\includegraphics[width=4.7cm]{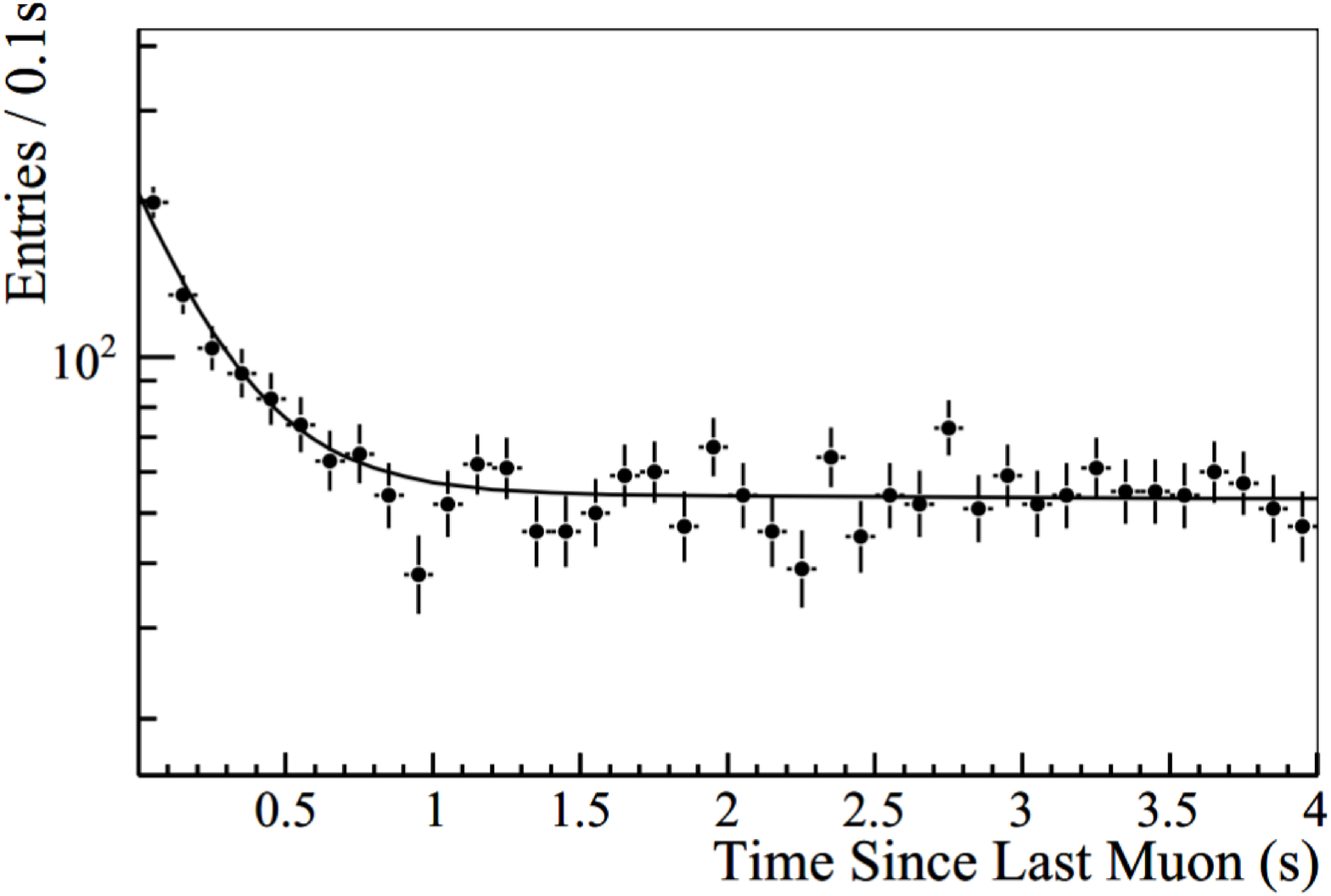}}
\vspace*{8pt}
\caption{Fitting method of $^8He/^9Li$ \label{li_fit}}
\end{minipage}
\end{figure}

The $^8He/^9Li$ isotopes produced by cosmic muons have substantial beta-neutron decay branching fractions (see Fig.\ref{li_pro}). The beta energy of the beta-neutron cascade overlaps the energy range of positron signals from neutrino events, mimicking the prompt signal, and the neutron emission forms the delayed signal.
\begin{equation}
f(t)=N_{Li+He}(R\cdot\lambda_{Li}\cdot e^{-\lambda_{Li}\cdot t}+(1-R)\cdot\lambda_{He}\cdot e^{-\lambda_{He}\cdot t})+N_{ibd}\cdot R_\mu \cdot e^{-R_\mu\cdot t}.
\label{lifunc}
\end{equation}
Fit the time since last muon distribution with Eq.\ref{lifunc} to measure $^8He/^9Li$ background\cite{WLJ}. Fig.\ref{li_fit} shows an example of fitting for $^8He/^9Li$ backgrounds.
\section{Results}
The muon induced backgrounds in six antineutrino detectors(AD) after muon rejection is listed in following table\cite{CPC}.

\begin{table}[!htb]
\tbl{Result of backgrounds in Daya Bay  Reactor Neutrino Experiment}
{\begin{tabular}{@{}ccc|c|ccc@{}}
  \toprule
   & AD1 & AD2 & AD3 & AD4 & AD5 & AD6 \\
   \colrule
  IBD rate(/day) & 662.47$\pm$3.00 & 670.87$\pm$3.01 & 613.53$\pm$2.69 & 77.57$\pm$0.85 & 76.62$\pm$0.85 & 74.97$\pm$0.84 \\

  Accidentals(/day) & 9.73$\pm$0.10 & 9.61$\pm$0.10 & 7.55$\pm$0.08 & 3.05$\pm$0.04 & 3.04$\pm$0.04 & 2.93$\pm$0.03 \\

  Fast neutron(/day)  & 0.77$\pm$0.24 & 0.77$\pm$0.20 & 0.58$\pm$0.33 & 0.05$\pm$0.02 & 0.05$\pm$0.02 & 0.05$\pm$0.02\\

  $^{8}$He/$^{9}$Li(/day) & \multicolumn{2}{c|}{2.9$\pm$1.5} & 2.0$\pm$1.1 & \multicolumn{3}{c}{0.22$\pm$0.12} \\
  \botrule
\end{tabular}\label{22}}
\end{table}

\section*{Acknowledgments}

Sincerely thanks to every contributor to this study at Daya Bay Collaboration. Thanks to everyone who helped me .


\end{document}